# It's all in the (Sub-)title? Expanding Signal Evaluation in Crowdfunding Research

Andrew Jay Isaak
*Heinrich-Heine University of Duesseldorf*, andrew.isaak@solbridge.ac.kr

Constantin von Selasinsky
*Heinrich-Heine University of Duesseldorf*, constantin.von.selasinsky@uni-duesseldorf.de

# IT'S ALL IN THE (SUB-)TITLE? EXPANDING SIGNAL EVALUATION IN CROWDFUNDING RESEARCH

*Research in Progress*

von Selasinsky, Constantin, Heinrich-Heine University of Duesseldorf, Düsseldorf, Germany, constantin.von.selasinsky@hhu.de

Isaak, Andrew, Heinrich-Heine University of Duesseldorf, Düsseldorf, Germany and SolBridge International School of Business, South Korea, andrew.isaak@hhu.de[*]

[*]Both authors contributed equally to this study

## Abstract

*Research on crowdfunding success that incorporates CATA (computer-aided text analysis) is quickly advancing to the big leagues (e.g., Parhankangas and Renko, 2017; Anglin et al., 2018; Moss et al., 2018) and is often theoretically based on information asymmetry, social capital, signaling or a combination thereof. Yet, current papers that explore crowdfunding success criteria fail to take advantage of the full breadth of signals available and only very few such papers examine technology projects. In this paper, we compare and contrast the strength of the entrepreneur's textual success signals to project backers within this category. Based on a random sample of 1,049 technology projects collected from Kickstarter, we evaluate textual information not only from project titles and descriptions but also from video subtitles. We find that incorporating subtitle information increases the variance explained by the respective models and therefore their predictive capability for funding success. By expanding the information landscape, our work advances the field and paves the way for more fine-grained studies of success signals in crowdfunding and therefore for an improved understanding of investor decision-making in the crowd.*

## 1 Introduction

Crowdfunding is an alternative type of project financing where a large and dispersed online audience contributes small financial amounts in exchange for tangible or intangible rewards. Crowdfunding differs from traditional seed finance and bank loans by attracting small investments from less sophisticated investors in a computer-mediated online setting. Crowdfunding research is emerging as a major area of entrepreneurial finance research with over 65 publications in top journals in the last 5 years. As such, the phenomenon has attracted the attention of scholars from different fields who defined the concept (e.g., Mollick, 2014), explored various factors affecting success rates (e.g., Agrawal et al., 2015; Courtney et al., 2017) and sought to understand the theoretical mechanisms behind the process and actions of the participants (e.g., Belleflamme et al., 2013; Xu et al., 2016) and the effect on industries (e.g., Gamble et al., 2017). Research on crowdfunding success that incorporates CATA (computer-aided text analysis) is quickly advancing to the big leagues (e.g., Parhankangas and Renko, 2017; Anglin et al., 2018; Moss et al., 2018) and is often theoretically based on information asymmetry, impression management or signaling (e.g., Ahlers et al., 2015). Signaling theory (Spence, 1978) posits that actors overcome information imbalances (asymmetries) by sharing information in the form of signals, which have to be both observable and costly to be effective (e.g., Connelly et al., 2011).

Current crowdfunding research elaborates on constructs to show the interplay and dynamics between signals and seeks to evaluate the effectiveness of various investor signals (e.g., visual and textual) across different settings (e.g., Allison et al., 2017; Scheaf et al., 2018). Yet, current papers exploring crowdfunding success criteria fail to take advantage of the full breadth of investor signals available for study. In this paper, we compare and contrast the strength of the entrepreneur's text-based signals to project backers while increasing the scope of information available for content analysis by utilizing transcribed video subtitles. The objective of our study is to answer the following research question: Among text-based investor signals in crowdfunding projects, what are the comparative signal strengths among linguistic constructs and what informational value does subtitle text contribute above and beyond the text listed in the project description?

Based on a random sample of crowdfunded technology projects, our study finds that incorporating subtitle information increases the variance explained by the respective regression models and therefore their predictive capability for funding success. Our paper contributes to linguistic research on crowdfunding (e.g., Davis et al., 2017), deception detection (Zhou et al., 2004) and artificial intelligence (Mairesse et al., 2007; Biyani et al., 2016). By expanding the information landscape, our work advances the field and paves the way for more fine-grained studies of success signals in crowdfunding and therefore for an improved understanding of investor decision-making in the crowd.

## 2 Theory and Hypotheses

Following uncertainty reduction theory, exchanging and collecting information on each other reduces uncertainty and allows one to predict others' attitudes and behaviors (Berger and Calabrese, 1974). In initial encounters, strangers follow specific verbal and nonverbal steps to create positive impressions on others, and to facilitate their judgments about people and situations. Since in crowdfunding, entrepreneurs typically have only one chance to make a first impression, successful crowdfunding campaigns are strongly determined by entrepreneurs' effective communication (Davis and Webb, 2012; Agrawal et al., 2015), which can mitigate the information asymmetry between entrepreneurs and investors in the crowd in a setting where the parties do not know each other, reducing the risk of a non-relevant investment (Block et al., 2018; Moss et al., 2018). Following from signaling theory (Spence, 1978; Connelly et al., 2011; Ahlers et al., 2015) effective communication on behalf of the entrepreneur in his or her campaign project descriptions or video pitches are signs of project quality, the entrepreneur's professionalism, trustworthiness, etc. These signals are perceived by backers in the crowd and their shape determines their effectiveness. A number of linguistic categories have been found to contain informational value for crowdfunding and crowdlending decisions, including textual complexity, affect and (in)formality (e.g., Allison et al., 2017; Zhang et al., 2017). Individuals' evaluations (e.g., decisions whether to back a crowdfunding project) can be shaped by peripheral cues (Crano and Prislin, 2006). For example, research has suggested that higher complexity in language use negatively effects readability since customers want convenience. Overly complex language can therefore inhibit purchase (or donation) intent (Block et al., 2018). Further, the nature of technology projects is to provide specific, measurable advancements and this domain is characterized by the use of concrete facts and figures (e.g., a drone that reaches an altitude of 500 meters or a car that reaches 0 to 60 in under 3 seconds) (Kickstarter, 2020). Low specificity has also been linked to deceptive communication (e.g., Burns and Moffitt, 2014) and it is clearly harder to make a good decision on investing in or supporting a new technology without precise information on its usage benefits. Therefore, we posit that:

> **H1**: Higher complexity and lower specificity of language in pitches and project descriptions negatively predict crowdfunding success.

The effectiveness of persuasion in communication may be influenced by the presence of affect in the message (Miniard et al., 1991) or even the tone or mood (Yang et al., 2006). Narratives frequently adopt an optimistic, positive tone in an effort to craft a likeable story (Martens et al., 2007), while negative emotionality should be counterproductive. Further, the use of positive tone increases the chance that individuals will be liked by others (Curtis and Miller, 1986) and results by (Allison et al., 2017) suggest that this also applies to crowdfunding pitches. As a consequence of computer-mediated communication, entrepreneurs have considerable control over communication and can take time to craft a convincing narrative by editing messages iteratively and are therefore likely to use higher expressivity in their writing. Technology entrepreneurs may even hire or partner with external public relations agencies, which are likely to craft persuasive pitches. Finally, herding has been linked to emotionality in online investment behavior (e.g., Zhang et al., 2017). Therefore, we expect that for technology projects:

> **H2**: Lower negative affect and higher expressivity of pitches and project descriptions positively predict crowdfunding success.

Professionalism in communication has been frequently linked to successful crowdfunding outcomes (e.g., Davis and Webb, 2012; Agrawal et al., 2015). Also, informal language can be highly indicative of spam (Biyani et al., 2016) and implies that entrepreneurs did not put in much time or effort into crafting their crowdfunding campaign. This seems particularly troublesome for technology projects. Therefore, we posit that:

> **H3**: Higher informality of pitches and project descriptions decreases crowdfunding success likelihood.

It has been found that higher uncertainty in language is indicative of deception in asynchronous computer-mediated communication (e.g., Zhou et al., 2004), while linguistic markers of certainty, such as "always" or "never," are strong indicators of truthfulness (Rubin et al., 2006; Levitan et al., 2018). Uncertainty reduction theory (Berger and Calabrese, 1974) suggests that there is a human drive to reduce uncertainty, to explain the world, and to render it predictable; in our context, crowd investors seek additional information related to assets to reduce their risk of making a false decision to support a project. At the same time, discrepancies in the entrepreneurs' language are likely to decrease investor confidence. Also, use of discrepancies in language have been found to be related to cognitive processes linked to negative emotions in artificial intelligence research on personality (Mairesse et al., 2007). Therefore, collectively we posit that:

> **H4**: Higher textual certainty and lower discrepancy in pitches and project descriptions positively predicts crowdfunding success.

In order to extract language-based information entailed in video pitches, these must first be transcribed. While manual transcription is time consuming, automatic transcription is rumored to be inaccurate (Ziman et al., 2018). In this study, we examine the informational value of the full narratives, i.e. project descriptions and the corresponding video pitches. While this approach allows for a full-fledged analysis of the informative power of language in crowdfunding campaigns, we argue that within the setting of technology *von Selasinsky and Isaak / Expanding Signal Evaluation in Crowdfunding*projects, project descriptions are a stronger predictor of funding success. Kickstarter assists with DIY-instructions for videos suggesting to project creators that "it doesn't have to be super slick" as long as the audience gets a feeling for the character of the project. However, videos inclusion is not compulsory. Rather, Kickstarter strongly emphasizes providing detailed information on the project page, such as personal information, manufacturing and budget plans and a schedule (Kickstarter, 2020). Further, projects in the technology category are heavy on specifications and functionality (Manning and Bejarano, 2017). While visual cues transport contextual information (e.g., age, attractiveness) technical terms are mostly transmitted in textual form (Scheaf et al., 2018). Therefore, we posit that:

**H5a**: Project descriptions are a stronger predictor of funding success than transcribed video pitches.

Using 60 manually transcribed video pitches as a benchmark, we compared the respective transcription accuracy on a word level (Ziman et al., 2018). With a hit rate of over 90%, Otter.ai (hereafter: Otter) outperformed YouTube (approximately 85%). We expect that the higher the accuracy of the transcription, the more information is captured. Hence, the outcome is closer to the message that was intended to be transmitted and more likely to be interpreted logically. Therefore, we suppose that YouTube subtitles have a lower explanatory power than their Otter counterparts and posit that:

**H5b**: Subtitle transcription quality positively impacts their predictive ability for crowdfunding success.

## 3 Data and Methodology

To answer our research question, we first drew a random sample of 1,099 US-based technology projects from the crowdfunding website Kickstarter in 2018. Technology projects were chosen because they are more likely to resemble traditional entrepreneurial ventures (Scheaf et al., 2018). The focus on a single category also improves project comparability. We removed 44 projects that were falsely categorized as US-based but where actually based in China and 6 projects for which their campaign data was missing (e.g., funding goal), resulting in a final dataset of 1,049 projects. The crowdfunding project video pitches were transcribed using the automated algorithms of the online services YouTube and Otter and merged with textual project descriptions and general project characteristics (e.g., funding goal, number of backers). After pilot testing the subtitling accuracy of various video platforms using a subsample of 60 projects, these platforms were selected for their high accuracy. Next, computer-aided text analysis (CATA) was conducted using the software LIWC (Pennebaker et al., 2015) together with the built-in content analytic dictionary for English language source material. Since our dependent variable funding success is binary, we next ran logistic regressions using the linguistic constructs (Kim et al., 2016) based on the project description and transcribed video pitches as independent variables (e.g., complexity, specificity, affect, expressivity, informality, time orientation, certainty and discrepancy). For the included linguistic categories that are operationalized below in Table 1, the internal item consistency (alpha) values reported by the authors of the software ranged between .70 and .84 (Pennebaker et al., 2015). We tested for autocorrelation and found that the variance inflation factors were all below 3. Detailed descriptive statistics are provided in the Appendix (Table 4).

| Construct | Definition | Selected Measure(s) | Examples |
|---|---|---|---|
| Complexity | The level of syntactical structures used by the sender | # of long words (> six letters) | conductivity, usefulness, innovation |
| | | # of dictionary words | picture, greeting, clockwork |
| Specificity | The degree to which the sender specifies facts for the conveyed information | # of numbers used | 3, 16, 2400 |
| Affect | The degree to which the sender describes or conveys personal emotions in his writing | Emotional tone index (Cohn et al., 2004) | |
| | | # of sadness words | crying, grief, sad |

.
*Twenty-Eighth European Conference on Information Systems (ECIS2020) – A Virtual AIS Conference.*

| Expressivity | The degree to which the sender colors his writing | # of modifiers (adjectives & adverbs) | free, long, short; very, really |
| --- | --- | --- | --- |
| | | # of perceptual words | look, heard, feeling |
| Informality | The degree to which the sender uses markers of informal language | # of informal words (assents, fillers, swear words, netspeak) | OK, yes; imean, youknow; f#ck, damn, sh#t; lol thx |
| Time orientation | The degree to which the sender is focused on the past, present or future | # of present focus words | today, is, now |
| Cognitive processes | The degree to which the sender shows conviction or uncertainty in writing | # of certainty words | always, never |
| | | # of discrepancy words | should, would |

*Table 1. Operationalization of Linguistic Constructs (for H1-H5)*

Based on the literature, we also control for the project funding goal, the number of projects a project creator launched on Kickstarter in the past, media use in terms of both images and pitch videos (e.g., Courtney et al., 2017), project duration, the number of project updates (e.g., Mollick, 2014), the amount pledged by backers, the number of awards offered to backers, team size and the perceived creativity of the project (see Table 2 below). For instance, Lagazio and Querci (2018) find a positive impact of large team size on the probability of campaign success.

| Variable | Description | Mean | S.D. | Min | Max |
| --- | --- | --- | --- | --- | --- |
| Funding Success | A dichotomous variable indicating if a project reached its goal | 0.255 | 0.436 | 0 | 1 |
| Funding Goal | The amount of USD a project aims to fund | 58479 | 209000 | 5000 | 5000000 |
| Projects Created | The count of past Kickstarter projects a project creator launched | 1.377 | 1.337 | 1 | 26 |
| Video | A dummy variable indicating if the project includes a video pitch | 0.824 | 0.381 | 0 | 1 |
| Picture | A dummy indicating use of additional visuals in the description | 12.871 | 18.147 | 0 | 131 |
| Duration | The number of days that the funding campaign ran on Kickstarter | 37.182 | 12.417 | 7 | 60.042 |
| Updates | The total amount of updates as a natural log | 4.66 | 10.226 | 0 | 211 |
| Pledged | The total sum of funds granted to the projects as a natural log | 6.39 | 3.743 | 0 | 14.96 |
| Reward level | The amount of different rewards a project offers to backers | 6.52 | 4.338 | 1 | 33 |
| Team | Coded "1" if several people collaborate on a project, "0" otherwise | 0.294 | 0.456 | 0 | 1 |
| Creativity | Describes the perceived creativity of the project, rated low to high | 1.751 | 0.748 | 1 | 3 |

*Table 2. Overview of Dependent Variable and Controls (n=1049)*

We interpret our results for hypotheses 1-4 based on the regression of our linguistic constructs on crowdfunding success (Model 2). First, we find that for technology projects, higher use of dictionary words is negatively related to crowdfunding success (-0.120, p<.001) in line with our hypothesis on complexity (**H1**), while the use of longer words is not significant. Similarly, we find that the higher use of numbers in text (our measure of specificity) positively predicts funding success as hypothesized (0.236, p=.001). Next, we find that sadness words, our measure of lower negative affect is indeed positively related with our dependent variable (-1.516, p=.003), while use of adjectives, adverbs and perceptual verbs (our measures of expressivity) are positively related to funding success (0.213, p=.003; 0.210, p<.021; 0.252, p<.001) as predicted (**H2**). Further, we find that higher use of informal language decreases the likelihood of crowdfunding success (-0.941, p<.001) as predicted (**H3**). Regarding the linguistic category cognitive processes, we find that higher certainty (0.528, p<.001) and lower discrepancy (-0.422, p=.007) in project descriptions and transcribed pitches predict crowdfunding success as predicted (**H4**).

| Model | 1 | 2 | 3 | 4 | 5 | 6 | 7 |
|---|---|---|---|---|---|---|---|
| Variable | Controls | Proj.Desc. | Controls& Proj.Desc. | Otter | Controls& Otter | YouTube | Controls& YouTube |
| goal | -0.000*** (0.000) | | -0.000*** (0.000) | | -0.000*** (0.000) | | -0.000*** (0.000) |
| created | -0.133 (0.088) | | -0.138 (0.103) | | -0.132 (0.111) | | -0.192* (0.108) |
| video | -0.453 (0.948) | | -1.031 (1.066) | | | | |
| picture | -0.003 (0.014) | | -0.009 (0.016) | | -0.002 (0.018) | | -0.022 (0.020) |
| duration | -0.070*** (0.024) | | -0.090*** (0.029) | | -0.111*** (0.036) | | -0.124*** (0.044) |
| updates | 0.189*** (0.055) | | 0.291*** (0.073) | | 0.335*** (0.082) | | -0.308*** (0.097) |
| log pledged | 2.488*** (0.324) | | 2.795*** (0.401) | | 2.972*** (0.532) | | 4.038*** (0.869) |
| reward levels | 0.079 (0.062) | | 0.139** (0.069) | | -0.005 (0.078) | | 0.027 (0.104) |
| team size | -0.17 (0.451) | | -0.011 (0.539) | | 0.192 (0.599) | | 0.565 (0.761) |
| creativity | 0.745** (0.364) | | 1.012** (0.460) | | 0.957* (0.495) | | 1.414** (0.642) |
| word count | | 0.001*** (0.000) | -0.000 (0.000) | 0.000 (0.000) | -0.003*** (0.001) | 0.000 (0.000) | -0.003 (0.002) |
| sixletter words | | -0.008 (0.023) | 0.192* (0.076) | -0.012 (0.020) | 0.071 (0.068) | -0.031 (0.021) | 0.1 (0.070) |
| dictionary | | -0.120*** (0.018) | 0.031 (0.058) | -0.095*** (0.017) | 0 (0.054) | -0.097*** (0.019) | 0.027 (0.058) |
| numbers | | 0.236*** (0.072) | 0.610*** (0.229) | 0.126* (0.068) | 0.488* (0.289) | 0.176** (0.075) | 0.820** (0.387) |
| tone | | 0.019*** (0.005) | 0.028* (0.017) | 0.012*** (0.004) | -0.005 (0.016) | 0.009* (0.005) | -0.041** (0.021) |
| sadness | | -1.516*** (0.517) | -1.511 (1.573) | -0.322 (0.322) | -0.719 (1.041) | -0.045 (0.319) | -0.366 (1.213) |
| adjectives | | 0.213*** (0.072) | -0.083 (0.210) | 0.187*** (0.046) | 0.157 (0.167) | 0.160*** (0.056) | 0.338 (0.256) |
| adverbs | | 0.210** (0.091) | 0.439* (0.250) | 0.06 (0.051) | 0.112 (0.181) | 0.083 (0.059) | 0.274 (0.229) |
| perceptual | | 0.252*** (0.056) | -0.2 (0.131) | 0.154*** (0.046) | -0.263** (0.132) | 0.025** (0.010) | 0.001 (0.038) |
| informal | | -0.941*** (0.176) | -1.302*** (0.438) | -0.161*** (0.053) | 0.101 (0.225) | -0.053 (0.101) | -0.052 (0.347) |
| certainty | | 0.528*** (0.132) | 0.358 (0.394) | 0.125 (0.077) | 0.145 (0.301) | 0.161 (0.104) | -0.231 (0.402) |
| discrepancy | | -0.422*** (0.157) | 0.199 (0.492) | 0.047 (0.101) | -0.332 (0.366) | -0.088 (0.113) | -0.134 (0.501) |
| Obs. | 1049 | 1049 | 1049 | 660 | 660 | 533 | 533 |
| Pseudo $R^2$ | 0.871 | 0.289 | 0.896 | 0.117 | 0.889 | 0.109 | 0.899 |
| Standard errors are in parenthesis, *** $p < 0.01$, ** $p < 0.05$, * $p < 0.1$ | | | | | | | |

*Table 3. Regression Results, Models 1-7 (DV: Funding Success)*

Finally, comparing across models, we find that the McFadden's R square for Model 2 (0.289), which regresses the linguistic constructs in project descriptions and funding success, is higher than that for Models 4 (0.117) and 6 (0.109), for which the transcribed video pitches from YouTube and Otter serve as a basis for the linguistic variables. Therefore, we can conclude that project descriptions are a stronger predictor of funding success than transcribed video pitches, as predicted (**H5a**). Based on our own experience and analysis of a subsample of professionally transcribed video pitches, Otter's machine-learning-based transcription algorithm is clearly superior to that of YouTube. We were therefore surprised to find that when comparing Model 3 with Models 5 and 7, which include our general project control variables, this increase in McFadden's R square holds only for Otter but not for YouTube. That is, the linguistic constructs based on the crowdfunding project descriptions together with the control variables explain more than the video pitches transcribed by Otter together with the control variables, but this is not the case when the same video pitches are transcribed by YouTube. Therefore, we must reject our hypothesis that subtitle transcription quality positively impacts their predictive ability for crowdfunding success (**H5b**).

## 4      Discussion and Implications

First, the results of the effects of our linguistic constructs on crowdfunding success in technology ventures (H1-H4) largely confirm our predictions and fall in line with previous literature on artificial intelligence (Mairesse et al., 2007; Biyani et al., 2016), deception detection (Zhou et al., 2004) and crowdfunding (Davis et al., 2017). Further, to the best of our knowledge, we provide the first direct evidence that project descriptions have higher predictive capability (and thus signal strength) for crowdfunding success than video subtitles and indeed provide additional informational value. At the same time, our finding that automatic transcription quality of video pitch subtitles by both YouTube and Otter provide close approximations of human coded transcription for the purposes of big data analytics is a valuable insight for the IS community and for linguistic research on crowdfunding and related fields. Methodologically, our results imply that the widely adopted practice in crowdfunding literature of analyzing only project descriptions is a good approximation of the full narrative, and for the IS community that researchers transcribing pitches are well served by using automated approaches, with little added value provided by manual transcription over AI for big data purposes. By expanding the information landscape regarding factors that determine a successful crowdfunding campaign, our work advances the field and paves the way for more fine-grained studies of success signals in crowdfunding and therefore for an improved understanding of venture funding decisions in the crowd. Enlarging the data available for CATA also reduces the risk of findings due to chance, helping to overcome a current weakness in crowdfunding research. Our study also has practical implications. First, project creators can incorporate the uncovered linguistic signals in the study to better craft their campaign narratives and improve their chances of funding success. Further, the study implies that subtitle quality is a viable and observable investor signal for campaign quality for both project backers and crowdfunding platforms at large. This could be used to better detect both high quality and fraudulent projects.

## 5      Limitations and Future Research

Our study has several limitations. First, our focus on technology projects means that our findings may not generalize beyond this category. Nonetheless, we would expect similar results for the category technology projects on similar US-based crowdfunding platforms such as Indiegogo. Future research can expand on our work by comparing and contrasting our results with those on other platforms and in other categories. Further, as this is research-in-progress, our analysis is preliminary and we have, for example, not yet examined the detailed effects of various interactions on our results.

# References

Agrawal, A., C. Catalini and A. Goldfarb (2015). "Crowdfunding: Geography, social networks, and the timing of investment decisions." *Journal of Economics & Management Strategy* 24(2), 253-274.

Ahlers, G. K., D. Cumming, C. Günther and D. Schweizer (2015). "Signaling in equity crowdfunding." *Entrepreneurship Theory and Practice* 39(4), 955-980.

Allison, T. H., B. C. Davis, J. W. Webb and J. C. Short (2017). "Persuasion in crowdfunding: An elaboration likelihood model of crowdfunding performance." *Journal of Business Venturing* 32(6), 707-725.

Anglin, A. H., J. C. Short, W. Drover, R. M. Stevenson, A. F. McKenny and T. H. Allison (2018). "The power of positivity? The influence of positive psychological capital language on crowdfunding performance." *Journal of Business Venturing* 33(4), 470-492.

Belleflamme, P., T. Lambert and A. Schwienbacher (2013). "Individual crowdfunding practices." *Venture Capital* 15(4), 313-333.

Berger, C. R. and R. J. Calabrese (1974). "Some explorations in initial interaction and beyond: Toward a developmental theory of interpersonal communication." *Human Communication Research* 1(2), 99-112.

Biyani, P., K. Tsioutsiouliklis and J. Blackmer (2016). ""8 Amazing Secrets for Getting More Clicks": Detecting Clickbaits in News Streams Using Article Informality." In: *Thirtieth AAAI Conference on Artificial Intelligence*.

Block, J., L. Hornuf and A. Moritz (2018). "Which updates during an equity crowdfunding campaign increase crowd participation?" *Small Business Economics* 50(1), 3-27.

Burns, M. B. and K. C. Moffitt (2014). "Automated deception detection of 911 call transcripts." *Security Informatics* 3(1), 8.

Connelly, B. L., S. T. Certo, R. D. Ireland and C. R. Reutzel (2011). "Signaling theory: A review and assessment." *Journal of Management* 37(1), 39-67.

Courtney, C., S. Dutta and Y. Li (2017). "Resolving information asymmetry: Signaling, endorsement, and crowdfunding success." *Entrepreneurship Theory and Practice* 41(2), 265-290.

Crano, W. D. and R. Prislin (2006). "Attitudes and persuasion." *Annual Review of Psychology* 57, 345-374.

Curtis, R. C. and K. Miller (1986). "Believing another likes or dislikes you: Behaviors making the beliefs come true." *Journal of Personality and Social Psychology* 51(2), 284.

Davis, B. C., K. M. Hmieleski, J. W. Webb and J. E. Coombs (2017). "Funders' positive affective reactions to entrepreneurs' crowdfunding pitches: The influence of perceived product creativity and entrepreneurial passion." *Journal of Business Venturing* 32(1), 90-106.

Davis, B. C. and J. W. Webb (2012). "Crowdfunding of entrepreneurial ventures: Getting the right combination of signals." *Frontiers of Entrepreneurship Research*.

Gamble, J. R., M. Brennan and R. McAdam (2017). "A rewarding experience? Exploring how crowdfunding is affecting music industry business models." *Journal of Business Research* 70, 25-36.

Kickstarter (2020). *Kickstarter Creator Handbook: Telling Your Story*. https://www.kickstarter.com/help/handbook/your_story (visited on Jan. 14, 2020).

Kim, P. H., M. Buffart and G. Croidieu (2016). "TMI: Signaling credible claims in crowdfunding campaign narratives." *Group & Organization Management* 41(6), 717-750.

Lagazio, C. and F. Querci (2018). "Exploring the multi-sided nature of crowdfunding campaign success." *Journal of Business Research* 90, 318-324.

Levitan, S. I., A. Maredia and J. Hirschberg (2018). "Linguistic cues to deception and perceived deception in interview dialogues." In: *Proceedings of the 2018 Conference of the North American Chapter of the Association for Computational Linguistics: Human Language Technologies, Volume 1 (Long Papers)*. 1941-1950.

Mairesse, F., M. A. Walker, M. R. Mehl and R. K. Moore (2007). "Using linguistic cues for the automatic recognition of personality in conversation and text." *Journal of Artificial Intelligence Research* 30, 457-500.

Manning, S. and T. A. Bejarano (2017). "Convincing the crowd: Entrepreneurial storytelling in crowdfunding campaigns." *Strategic Organization* 15(2), 194-219.

Martens, M. L., J. E. Jennings and P. D. Jennings (2007). "Do the stories they tell get them the money they need? The role of entrepreneurial narratives in resource acquisition." *Academy of Management Journal* 50(5), 1107-1132.
Miniard, P. W., S. Bhatla, K. R. Lord, P. R. Dickson and H. R. Unnava (1991). "Picture-based persuasion processes and the moderating role of involvement." *Journal of Consumer Research* 18(1), 92-107.
Mollick, E. (2014). "The dynamics of crowdfunding: An exploratory study." *Journal of Business Venturing* 29(1), 1-16.
Moss, T. W., M. Renko, E. Block and M. Meyskens (2018). "Funding the story of hybrid ventures: Crowdfunder lending preferences and linguistic hybridity." *Journal of Business Venturing* 33(5), 643-659.
Parhankangas, A. and M. Renko (2017). "Linguistic style and crowdfunding success among social and commercial entrepreneurs." *Journal of Business Venturing* 32(2), 215-236.
Pennebaker, J. W., R. L. Boyd, K. Jordan and K. Blackburn (2015). *The Development and Psychometric Properties of LIWC2015*.
Rubin, V. L., E. D. Liddy and N. Kando (2006). Certainty identification in texts: Categorization model and manual tagging results. In: (Eds.), *Computing Attitude and Affect in Text: Theory and Applications*, p. 61-76: Springer.
Scheaf, D. J., B. C. Davis, J. W. Webb, J. E. Coombs, J. Borns and G. Holloway (2018). "Signals' flexibility and interaction with visual cues: Insights from crowdfunding." *Journal of Business Venturing* 33(6), 720-741.
Spence, M. (1978). Job market signaling. In: (Eds.), *Uncertainty in Economics*, p. 281-306: Elsevier.
Xu, B., H. Zheng, Y. Xu and T. Wang (2016). "Configurational paths to sponsor satisfaction in crowdfunding." *Journal of Business Research* 69(2), 915-927.
Yang, S. C., W. C. Hung, K. Sung and C. K. Farn (2006). "Investigating initial trust toward e-tailers from the elaboration likelihood model perspective." *Psychology & Marketing* 23(5), 429-445.
Zhang, S., R. C.-W. Kwok and Z. Liu (2017). "Reward-based crowdfunding: a study of online investment behaviors among China funders." In: *2017 2nd ACSS International Conference on the Social Sciences and Teaching Research (ACSS-SSTR 2017)*. Singapore Management and Sports Science Institute. 9-15.
Zhou, L., J. K. Burgoon, J. F. Nunamaker and D. Twitchell (2004). "Automating linguistics-based cues for detecting deception in text-based asynchronous computer-mediated communications." *Group Decision and Negotiation* 13(1), 81-106.
Ziman, K., A. C. Heusser, P. C. Fitzpatrick, C. E. Field and J. R. Manning (2018). "Is automatic speech-to-text transcription ready for use in psychological experiments?" *Behavior Research Methods* 50(6), 2597-2605.

## Appendix

| Variable | Obs. | Mean | S.D. | Min. | Max. |
| --- | --- | --- | --- | --- | --- |
| word count | 1049/660/533 | 852.95/391.548/367.326 | 692.168/410.039/286.815 | 6/6/1 | 5395/5008/2813 |
| sixltr words | 1049/660/533 | 24.486/19.157/18.604 | 4.795/5.555/6.921 | 8.51/1.56/0 | 46.26/40.97/50 |
| dictionary | 1049/660/533 | 79.201/81.711/83.8 | 6.787/8.118/7.568 | 45.96/33.33/50 | 95/95.53/100 |
| numbers | 1049/660/533 | 2.042/1.662/1.629 | 1.382/1.426/1.374 | 0/0/0 | 11.99/13.14/9.43 |
| tone | 1049/660/533 | 71.543/72.378/71.168 | 20.955/24.472/25.322 | 4/2.18/2.39 | 99/99/100 |
| sadness | 1049/660/533 | .162/.163/.168 | .268/.358/.368 | 0/0/0 | 3.45/4.95/4.57 |
| adjectives | 1049/660/533 | 4.414/4.536/4.387 | 1.397/2.13/2.086 | 0/0/0 | 9.96/18.52/13.16 |
| adverbs | 1049/660/533 | 3.221/4.419/4.26 | 1.239/2.163/2.253 | 0/0/0 | 8.51/12.16/14.29 |
| perceptual | 1049/660/533 | 2.163/2.659/7.024 | 1.538/2.006/17.588 | 0/0/0 | 11.36/20/100 |
| informal | 1049/660/533 | .777/1.547/.819 | .914/2.759/1.153 | 0/.16/0 | 9.68/33.33/11.48 |
| certainty | 1049/660/533 | 1.375/1.481/1.448 | .733/1.151/1.031 | 0/0/0 | 4.97/16.67/6.56 |
| discrepancy | 1049/660/533 | 1.259/1.27/1.29 | .76/.997/1.07 | 0/0/0 | 5.81/7.67/10 |

*Table 4. Summary Statistics, Linguistic Constructs (Project Description/Otter Subtitles/YouTube Subtitles)*